\documentclass{sig-alternate}

\begin{document}
%
% --- Author Metadata here ---
\conferenceinfo{The 5th SNA-KDD Workshop'11}{(SNA-KDD'11), August 21, 2011, San Diego CA USA.}
%\CopyrightYear{2007} % Allows default copyright year (20XX) to be over-ridden - IF NEED BE.
%\crdata{0-12345-67-8/90/01}  % Allows default copyright data (0-89791-88-6/97/05) to be over-ridden - IF NEED BE.
% --- End of Author Metadata ---

\title{What Trends in Chinese Social Media}

%
% You need the command \numberofauthors to handle the 'placement
% and alignment' of the authors beneath the title.
%
% For aesthetic reasons, we recommend 'three authors at a time'
% i.e. three 'name/affiliation blocks' be placed beneath the title.
%
% NOTE: You are NOT restricted in how many 'rows' of
% "name/affiliations" may appear. We just ask that you restrict
% the number of 'columns' to three.
%
% Because of the available 'opening page real-estate'
% we ask you to refrain from putting more than six authors
% (two rows with three columns) beneath the article title.
% More than six makes the first-page appear very cluttered indeed.
%
% Use the \alignauthor commands to handle the names
% and affiliations for an 'aesthetic maximum' of six authors.
% Add names, affiliations, addresses for
% the seventh etc. author(s) as the argument for the
% \additionalauthors command.
% These 'additional authors' will be output/set for you
% without further effort on your part as the last section in
% the body of your article BEFORE References or any Appendices.

\numberofauthors{3} %  in this sample file, there are a *total*
% of EIGHT authors. SIX appear on the 'first-page' (for formatting
% reasons) and the remaining two appear in the \additionalauthors section.
%
\author{
% You can go ahead and credit any number of authors here,
% e.g. one 'row of three' or two rows (consisting of one row of three
% and a second row of one, two or three).
%
% The command \alignauthor (no curly braces needed) should
% precede each author name, affiliation/snail-mail address and
% e-mail address. Additionally, tag each line of
% affiliation/address with \affaddr, and tag the
% e-mail address with \email.
%
% 1st. author
\alignauthor
Louis Yu\\
       \affaddr{Social Computing Lab}\\
	\affaddr{HP Labs}\\
	\affaddr{Palo Alto, California, USA}\\
       \email{louis.yu@hp.com}
% 2nd. author
\alignauthor
Sitaram Asur\\
      \affaddr{Social Computing Lab}\\
	\affaddr{HP Labs}\\
	\affaddr{Palo Alto, California, USA}\\
       \email{sitaram.asur@hp.com}\\
% 3rd. author
\alignauthor Bernardo A. Huberman\\
     \affaddr{Social Computing Lab}\\
	\affaddr{HP Labs}\\
	\affaddr{Palo Alto, California, USA}\\
       \email{bernardo.huberman@hp.com}
}

%\date{22 June 2011}

\maketitle

\begin{abstract}
There has been a tremendous rise in the growth of online social networks all over the world in recent times. While some networks like Twitter and Facebook have been well documented, the popular Chinese microblogging social network Sina Weibo has not been studied. In this work, we examine the key topics that trend on Sina Weibo and contrast them with our observations on Twitter.  We find that there is a vast difference in the content shared in China, when compared to a global social network such as Twitter. In China, the trends are created almost entirely due to retweets of media content such as jokes, images and videos, whereas on Twitter, the trends tend to have more to do with current global events and news stories.
\end{abstract}

% A category with the (minimum) three required fields
%A category including the fourth, optional field follows...



\section{Introduction}

Social networks have made tremendous impact on online computing, by providing users opportunities to connect with others and generate enormous content on a daily basis. The enormous user participation in these social networks is reflected in the
incessant number of discussions, images, videos, news and conversations that are constantly posted in social sites.
Popular networks such as Facebook and Twitter are well-known globally and contain several hundreds of millions of users all over the world. On the other hand, Sina Weibo is a popular microblogging network in China which contains millions of users, almost all of whom are located in China and post in the Chinese language.  

In China, online social networks have become a major platform for the youth to gather information and to make friends with like-minded individuals \cite{Jin}. In this regard, a major point of interest is to examine the information that is propagated and the key trend-setters for this medium. There has been a lot of prior research done on the adaptation of influence and evolution of trends in Western online social networks \cite{Asur2011} \cite{Kempe05influentialnodes}   \cite{Leskovec05patternsof}. But, in contrast, Chinese social media has not been well-studied.

In this paper, we analyze the evolution of Sina Weibo and provide the first known in-depth study of trending topics on a Chinese online microblogging social network. Our goal is to discover important factors that determine popularity and influence in the context of Chinese social media. To compare, we contrast them with corresponding ones from Western social media (Twitter). We put emphasis on examining  how trends are formed and what kind of sources dominate the topics of discussion in Chinese social media.

First, we identify and collect the topics that are popular on Sina Weibo over time. For each of these trending topics, we analyze the characteristics of the users and the corresponding tweets that are responsible for creating trends. We believe that this will give us a strong insight into the processes that govern social influence and adoption in China. To perform a comparison, we use similar trending topic data from Twitter.
 
Our key findings are as follows. We observe that there are vast differences between the content that is shared on Sina Weibo when compared to Twitter. In China, people tend to use Sina Weibo to share jokes, images and videos and a significantly large percentage of posts are retweets. The trends that are formed are almost entirely due to the repeated retweets of such media content. This is contrary to what we observe on Twitter, where the trending topics have more to do with current events and the effect of retweets is not as large. We also observe that there are more unverified accounts among the top 100 trend-setters on Sina Weibo than on Twitter and  most of the unverified accounts feature discussion forums for user-contributed jokes, images and videos.

\section{Background and Related Work}

\subsection{Degree Distributions}

%Many experiments have been conducted for studying the structure of the web \cite{Broder} \cite{Albert} \cite{WebGraph}.   One important property of the web is that it follows the power law in the in-degree and out-degree distribution of nodes \cite{Kumar-Trawling} \cite{Barabasi_1999} \cite{Albert} \cite{WebGraph}.  Researchers have taken snapshots of the web at different times and a recurring finding is that the fraction of web pages with degree $i$ is proportional to $1/i^{\alpha}$ for some constant $\alpha$.       

There have been many experiments conducted for studying the structure of online social networks. In one comprehensive study, Mislove et al. \cite{Mislove} have presented a large-scale measurement study of online social networks such as Orkut, YouTube, and Flickr. Their results show that online social networks follow the power-law in the in-degree and out-degree distributions of user nodes. In other work, Kumar et al. \cite{Kumar} examine the linking structure of Flickr and Yahoo!360 and report similar findings.  

\subsection{Social Influence Studies}

For many years the structure of various  offline social networks has been studied by sociologists (see \cite{Jamali} \cite{Mislove} \cite{Buchanan} for surveys).  Researchers have also analyzed the structure of various Chinese offline social networks \cite{StrongTie} \cite{WorkControl} \cite{768262} \cite{Guanxi3} \cite{Carrington}.  
 In social network analysis, \textit{social influence} refers to the concept of people modifying their behavior to bring them closer  to the behavior of their friends. 
%In a social-affiliation network consists of nodes representing individuals,  links representing friendships, and nodes representing \textit{foci}: ``social, psychological, legal, or physical entities around which joint activities are organized (e.g., workplace, social groups) \cite{mcpherson2001birds}'',  if $A$ and $B$ are friends, and $F$ is a focus that $A$ participate in.  Over time, $B$ can participate in the same focus due to $A$'s involvement, this is called a \textit{membership closure}\cite{mcpherson2001birds}.

Social influence has been studied in a vast array of social networks involving various foci such as interests and personal habits \cite{mcpherson2001birds} \cite{Goyal} \cite{Tang}. Xu et al. \cite{Xu} have looked at the adaptation of aggressive behaviors in a social network of kindergarden children in China.  As a method of controlling aggression, teachers in China tend to put aggressive children in a peer group with non-aggressive children.  Xu et al. \cite{Xu} have found that over time friendships can be formed between aggressive children and non-agressive children. For the aggressive children who are group members, intra-group friendships moderated their aggressive behavior.  

%\subsection{Social Influence in Online Social Networks}
Agarwal et al. ~\cite{Agarwal2008Identifying} have
examined the problem of identifying influential bloggers in the blogosphere.
They discovered that the most influential bloggers were not necessarily the most
active. Backstrom et al. \cite{Backstrom} have examined the characteristics of \textit{membership closure} in LiveJournal.   and Crandall et al. \cite{Crandall}, the adaptation of influences between editors of Wikipedia articles. 
On Twitter, Cha et al.~\cite{Cha2010Measuring} have performed a comparison
of three different measures of influence - indegree, retweets and user mentions.
Based on this, they hypothesized that the number of followers may not a good
measure of influence. This was corroborated by Romero and others~\cite{Romero2011} who presented 
a novel influence measure that takes into account the passivity of the audience in the social network. They measured retweets 
on Twitter and found that passivity was a major factor when it came to forwarding. They also demonstrated with empirical evidence that the number of followers is a poor measure of influence.

There are only a few studies of social influence in Chinese online social networks. 
Jin \cite{Jin} has studied the Chinese online Bulletin Board Systems (BBS), and provided observations on the structure and interface of Chinese BBS and the  behavioral patterns of its users.  Xin \cite{Xin} has conducted a survey on BBS's influence on the  University students in China and their behavior on Chinese BBS.  Yu et al.  \cite{yu} has looked at the adaptation of interests such as books, movies, music, events and discussion groups on Douban, an online social network frequently used by the youth in China. Douban provides users with review and recommendation services for movies, books, music and events. It is also the largest online media database and one of the largest online communities in China.

\subsection{Trends on Twitter}
There are various studies on trends on Twitter \cite{Huberman} \cite{Kwak} \cite{Mathioudakis} \cite{Wu2}.  Recently, Asur and others~\cite{Asur2011} have examined the growth and persistence of trending topics on Twitter. They discovered that traditional media sources are important in causing trends on twitter. Many of the top retweeted articles that formed trends on Twitter were found to arise from news sources such as the New York Times. In this work, we evaluate how the trending topics in China relate to the news media.

\subsection{The Internet in China}

The development of the Internet industry in China over the past decade has been impressive. According to a survey from  the China Internet Network Information Center (CNNIC), by July 2008, the number of Internet users in China has reached 253 million, surpassing the U.S. as the world's  largest Internet market \cite{Statistic-general}. Furthermore, the number of Internet users in China as of 2010 was reported to be 420 million.

Despite this, the fractional Internet penetration rate in China is still low.  The 2010 survey by CNNIC on the Internet development in China \cite{Statistic-rural} reports that the Internet penetration rate in the rural areas of  China is on average $5.1\%$. In contrast,  the  Internet penetration rate in the urban cities of China is on average $21.6\%$. In metropolitan cities such as Beijing and Shanghai, the Internet penetration rate has reached over $45\%$, with Beijing being $46.4\%$ and  Shanghai  being $45.8\%$ \cite{Statistic-rural}. 

\begin{figure} [ht]
\centering
\includegraphics  [width=80mm, height=50mm]{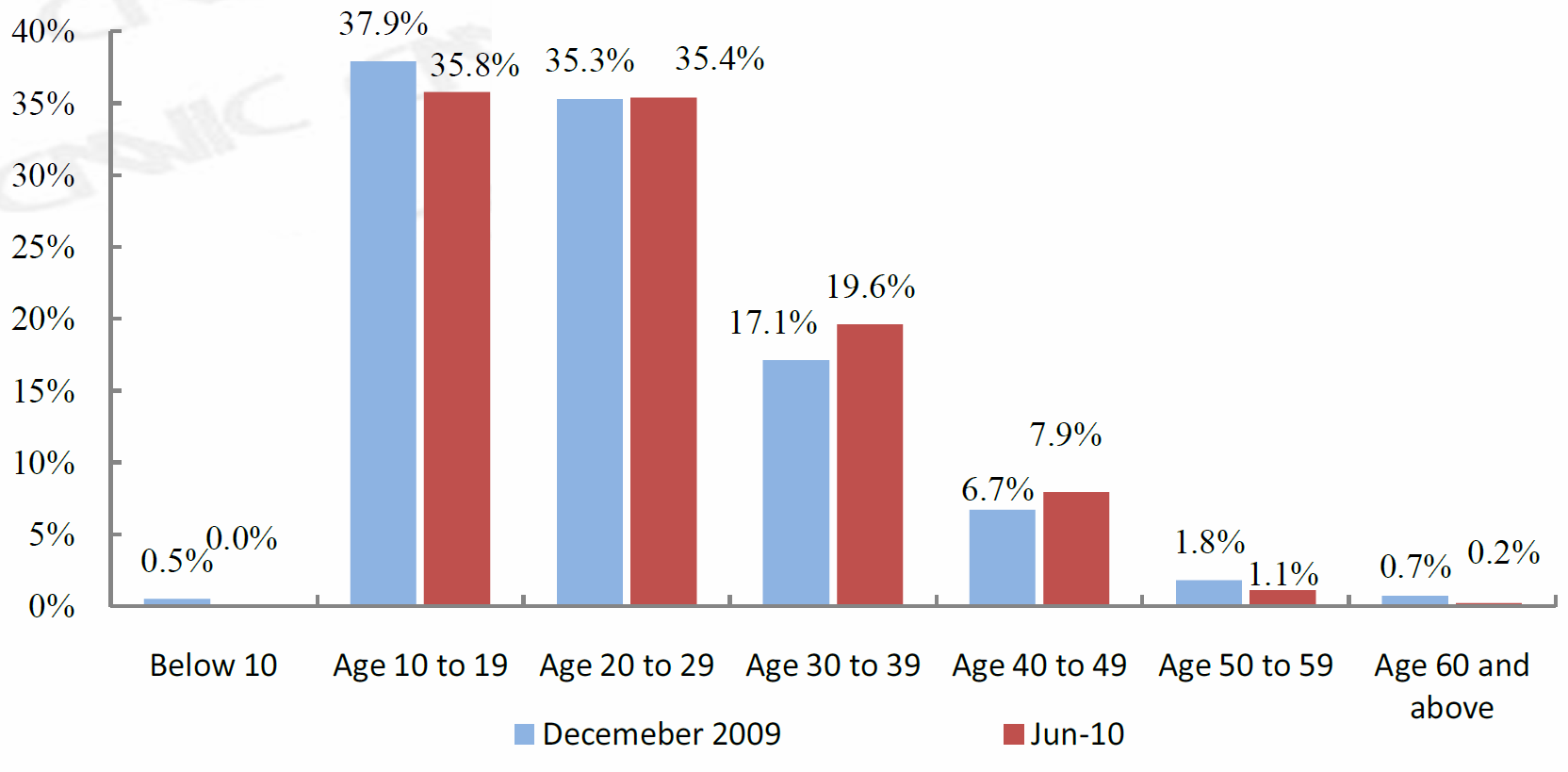}
\caption{ Age Distribution of Internet Users in China } \label{Age}
\end{figure}

\begin{figure} [ht]
\centering
\includegraphics [width=80mm, height=80mm]{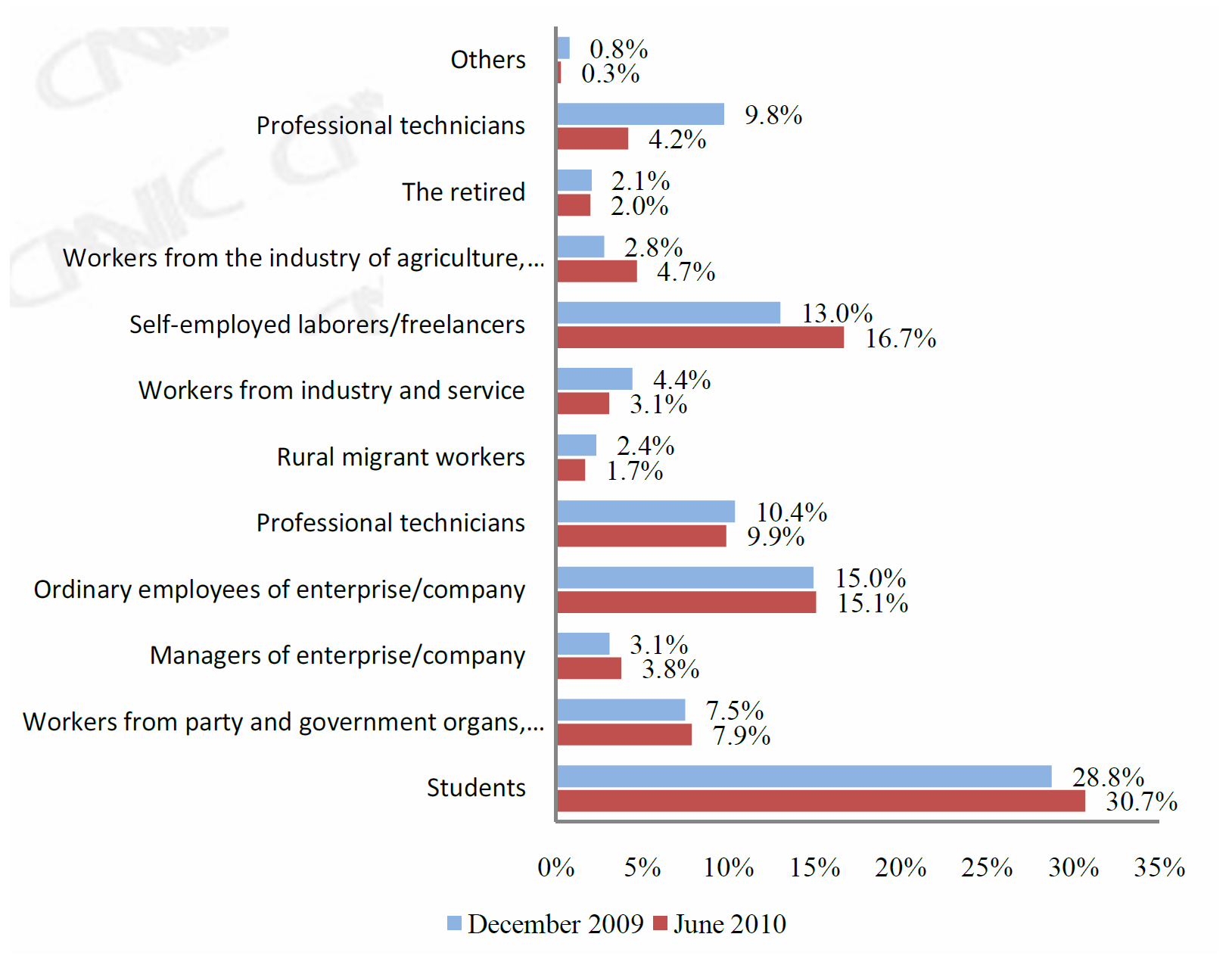}
\caption{ The Occupation Distribution of Internet Users in China } \label{Occupation}
\end{figure}

According to the survey by CNNIC in 2010 \cite{Statistic-general},  China's cyberspace is dominated by urban students between the age of 18--30 (see Figure \ref{Age} and Figure \ref{Occupation}, taken from \cite{Statistic-general}).  

The Government plays an important role in fostering the advance of the Internet industry in China.   Tai \cite{Tai} points out the four major stages of Internet development in China, ``with each period reflecting a substantial change not only in technological progress and application, but also in the Government's approach to and apparent perception of the Internet.'' 

\begin{enumerate}

\item The first phase was between 1986--1992, when Internet  applications were limited to the use of emails among a handful of computer research labs in China. 

\item The second phase was between 1992--1995, the Chinese Government proposed several large scale network projects and built a national information network infrastructure. 

\item The third phase was between 1995--1997. The Chinese Government stepped up its effort in building the information network infrastructure, hoping that the IT industry would yield significant benefits to the nation's economy. Meanwhile, the Government started to  implement a variety of technological and policy control mechanisms to regulate the safe flow of the information on the Internet. 
  
\item The fourth phase started from 1998 and continues to the present, during which time the Internet has become a powerful  medium in  the Chinese society. 

\end{enumerate}

 According to  \textit{The Internet in China} \footnote{``The Internet in China'' by the Information Office of the State Council of the People's Republic of China is available at \textit{http://www.scio.gov.cn/zxbd/wz/201006/t667385.htm}}  released by  the Information Office of the State Council of China: 

 \begin{quote}  The Chinese government attaches great importance to protecting the safe flow of Internet information, actively guides people to manage websites in accordance with the law and use the Internet in a wholesome and correct way. \end{quote}

\subsection{Chinese Online Social Networks}

 Online social networks are a major part of the Chinese Internet culture \cite{Jin}.  Netizens\footnote{A netizen is a person actively involved in online communities \cite{Netizen}.}  in China organize themselves using  forums, discussion groups, blogs, and social networking platforms to engage in  activities such as exchanging viewpoints and sharing information \cite{Jin}.  According to \textit{The Internet in China}:
 
 \begin{quote} 
 
 Vigorous online ideas exchange is a major characteristic of China's Internet development, and the huge quantity of BBS posts and blog articles is far beyond that of any other country. China's websites attach great importance to providing netizens with opinion expression services, with over 80\% of them providing electronic bulletin service. In China, there are over a million BBSs and some 220 million bloggers. According to a sample survey, each day people post over three million messages via BBS, news commentary sites, blogs, etc., and over 66\% of Chinese netizens frequently place postings to discuss various topics, and to fully express their opinions and represent their interests. The new applications and services on the Internet have provided a broader scope for people to express their opinions. The newly emerging online services, including blog, microblog, video sharing and social networking websites are developing rapidly in China and provide greater convenience for Chinese citizens to communicate online. Actively participating in online information communication and content creation, netizens have greatly enriched Internet information and content.
  \end{quote}

\section{Sina Weibo}

From the above motivation, we think it is interesting to look at how trends start and evolve in various Chinese online social networks and to analyze the characteristics of trend-setters determining if they represent Government organizations, commercial organizations, the media, or individuals. We choose to analyze the characteristics of trends and trend-setters on Sina Weibo.  Sina Weibo was  launched by the Sina corporation, China's biggest web portal, in August 2009.  On July 2009, the Chinese Government  blocked the access to Twitter and Fanfou, the then leading Twitter clone, in China.  Internet companies such as  Sina and Tencent started offering microblog services to their users in mainland China.  According to the  Sina corporation annual report \footnote{The Sina corporation annual report 2011 is available (in Chinese) at \textit{http://tech.sina.com.cn/i/2010-11-16/10314870771.shtml}}, the Weibo microblog now has more than 50 million active users per day, and 10 million newly registered users per month.

While both Twitter and Sina Weibo enable users to post messages of up to 140 characters, there are some differences in terms of the functionalities offered. We give a brief introduction of Sina Weibo's interface and functionalities. 

\subsection {User Profiles}

A user profile on Sina Weibo displays the user's name, a brief description of the user, the number of followers and followees the user has, and the number of tweets the user made. A user profile also displays the user's recent tweets and retweets. 

Similar to Twitter, there are two types of user accounts on Sina Weibo, regular user accounts and verified user accounts.  A verified user account typically represents a well known public figure or organization in China.  Sina has reported in the annual report that it has more than 60,000 verified accounts consisting of celebrities, sports stars, well known organizations (both Government and commercial) and other VIPs.  

%Figure \ref{verified} illustrates an example of a verified user account. 
% \begin{figure} [ht]
%\centering
%\includegraphics[width=80mm, height=90mm] {verified.png}
%\caption{ A Verified Account on Sina Weibo } \label{verified}
%\end{figure} 
%
%

\subsection{The Content of Tweets on Sina Weibo}

\begin{figure} [ht]
\centering
\includegraphics[width=80mm, height=90mm] {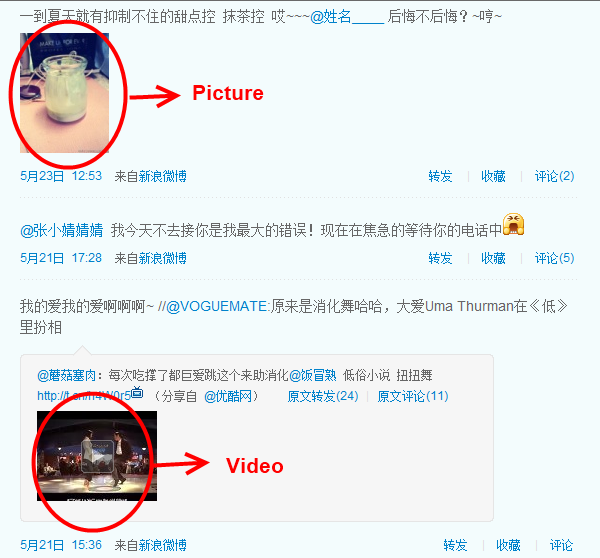}
\caption{ An Example of Embedded Videos and Pictures (Translations of the Tweets Omitted) } \label{video}
\end{figure}

There is an important difference in the content of tweets between Sina Weibo and Twitter.
While Twitter users can post tweets consisting of text and links,  Sina Weibo users can post messages containing text, pictures, videos and links. Figure \ref{video} illustrates some messages with embedded pictures and videos on Sina Weibo.

\subsection{Retweets and Comments}

\begin{figure} [ht]
\centering
\includegraphics[width=80mm, height=50mm] {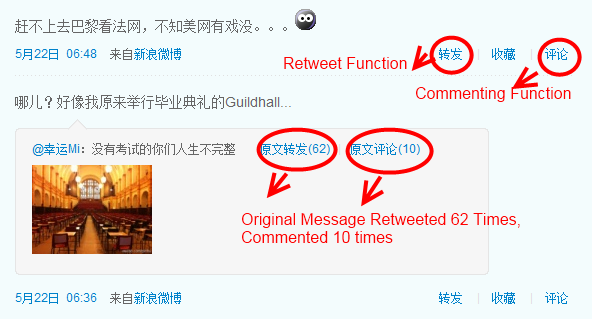}
\caption{ An Example of Comments and Retweets  (Translations of the Tweets Omitted) } \label{retweet}
\end{figure}

Twitter users can address tweets to other users and can mention others in their tweets [13].  A  common practice on Twitter is ``retweeting'',  or rebroadcasting someone 
else's messages to one's followers.  The equivalent of a retweet on Sina Weibo is instead shown as two two amalgamated entries: the original entry and the current user's actual entry which is a commentary on the original entry (see Figure \ref{retweet}).

Sina Weibo also has a functionality absent from Twitter: the comment. When a Weibo user makes a comment, it is not rebroadcasted to the user's followers. Instead, it can only be accessed under the original message. 

Figure \ref{retweet} illustrates two example tweets on Sina Weibo. The first is an original tweet made by a user, we can see that the retweeting and commenting buttons are listed under the tweet.  The second is a retweet, we can see that the original message is retweeted 62 times and commented 10 times by other users.  

\subsection{Trending keywords}

\begin{figure} [ht]
\centering
\includegraphics[width=80mm, height=90mm] {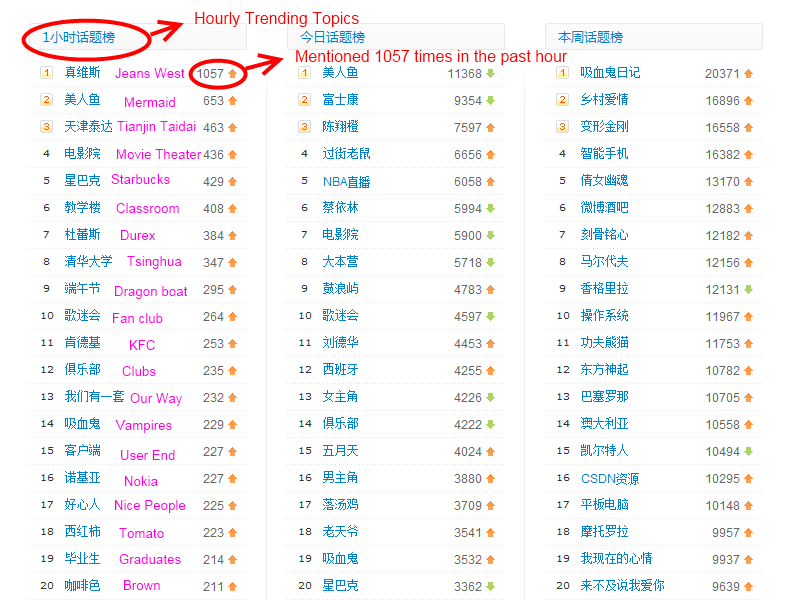}
\caption{ The List of Hourly Trending Keywords  (with Translations)} \label{Trends}
\end{figure}

Sina Weibo offers a list of 50 keywords that appeared most frequently in users' tweets over the past hour. They are ranked according to the frequency of appearances. Figure \ref{Trends} illustrates the list of hourly trending keywords (with translations). This is similar to Twitter, which also presents a constantly updated list of trending topics, which are keywords that are most frequently used in tweets over that period.

We monitored the list of hourly trending keywords every hour for 30 days and retrieved every new keywords appeared in the list. We retrieved in total 4411 new trending keywords over the 30 days observation period.

To compare with Twitter, we obtained 16.32 million tweets on 3361 different trending topics over 40 days using the Twitter Search API.

\section{Experiments and Results}

\begin{table*}[ht]
\caption{Top 20 Retweeted Users in At Least 10 Trending Topics} \label{Trend_Setter_Retweet}
\centering 
\begin{tabular}{|c|c|c|c|c|c|c|c|}
	\hline
	&ID &  Author Description (Translated)   & Verified Account &  Retweets  & Tweets & Topics   & Retweet-Ratio  \\
	\hline
 1&1757128873  & Urban Fashion Magazine & Yes & 1194999 &  37 & 12  & 99583.25\\
2& 1643830957 &  Fashion Brand VANCL  &   Yes &   849404 & 21  & 13  &  65338.77 \\
3&1670645393 & Online Travel Magazine &  Yes & 127737 & 123 & 21 & 57987.48\\
4&1992523932 & Gourmet Factory & No & 553586 & 86 & 12 & 46132.17\\
5&1735618041 & Horoscopes & No & 1545955 & 101 & 38 & 40683.13\\
 6&1644395354 & Silly Jokes& No & 3210130 & 258 & 81&39631.23\\
 7& 1843443790 & Good Movies & No & 1497968	& 140 & 38  &39420.21\\
8& 1644572034 & Wonderful Quotes	& No &	602528	&39& 17& 35442.82\\
9&1674242970&	Global Music &	No&	697308	&116	&22	& 31695.81\\
10&1713926427&	Funny Jokes Countdown &No&	3667566 &	438&	121	&30310.46\\
11&1657430300	& Creative Ideas 	& No & 	742178	& 111	& 25	& 29687.12\\
12&1195230310	& Famous Chinese singer & Yes & 	284600	& 25	 & 10 &	28460\\
13&1750903687	& Good Music	&No&	323022	& 52	& 12	& 26918.5\\
14&1757353251	& Movie Factory	& No & 	1509003	&230	&59	&25576.32\\
15&1644570320	&Strange Stories	&No &	1668910	&250	&66	&25286.52\\
16&1802393212	&Beautiful Pictures		&No&	435312&	33&	18&	24184\\
17&1920061532	&Global Music		&No	&432444	&65	&18	&24024.67\\
18&1644574352	&Female Fashion		&No	&809440	&87	&34	&23807.06\\
19&1780417033	&Useful Tips &	No&	735070&	153&	31&	23711.94\\
20&1644394154	&Funny Quizzes 		&No	&589477&	77&	25&	23579.08\\
	\hline	
\end{tabular}
\end{table*}
 \begin{figure*}
\centering
\includegraphics[width=170mm, height=70mm] {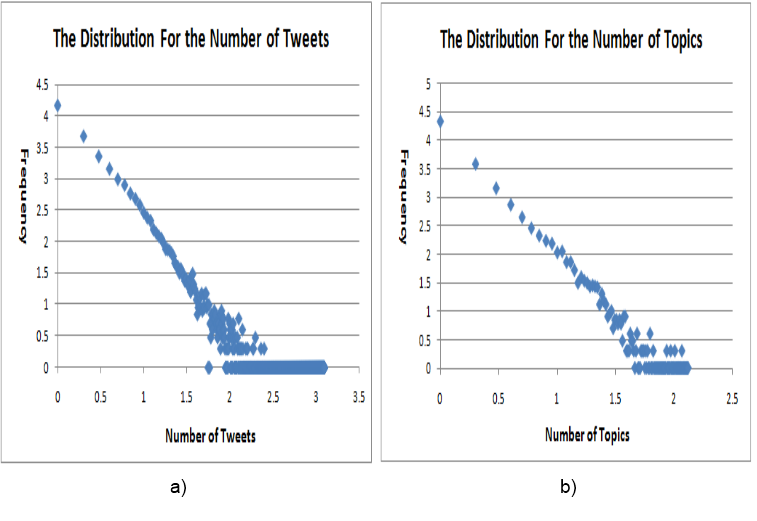}
\caption{ The Distribution for the Number of Tweets and the Number of Topics } \label{distribution}
\end{figure*} 

First, we calculated the distributions for the number of tweets and topics in our dataset. Figure \ref{distribution}  a) illustrates the distributions for the number of users (Y-axies) with a certain number of tweets (X-axis) in our list of trending topics; Figure \ref{distribution}  b) illustrates the distribution for the number of users (Y-axis) whose tweets appear in numbers of topics (X-axis). 
%users with a certain number of tweets in our list of trending topics, the distribution for the number of topics users' tweets %appeared in, and the distribution for the number of time users' tweets are retweeted. 
As we can observe from the figure, both these distributions follow the power law. 

\subsection{Trend-setters on Sina Weibo}

One of the main forms of information propagation in social networks such as Twitter and Sina is through retweets. When people find a tweet interesting either due to the content or the source, they forward it to their followers. 

For every new trending keyword we retrieved the most retweeted tweets in the past hour and compiled a list of most retweeted users.  Table \ref{Trend_Setter_Retweet} illustrates the top 20 most retweeted authors appearing in at least 10 trending topics each. We define an author's retweet ratio as the number of times the authors' tweets are retweeted divided by the number of trending topics these tweets appeared in. 

For each author we have included the ID of the user account (ID), a brief translation of the description of the authors (author description), whether it is a verified account, the number of tweets the author made in the trending topics (tweets), the number of times these tweets are retweeted, the number of topics the authors' tweets appeared in (topics), and finally, the influential authors are ranked according to their retweet ratios.  

\subsubsection{Authors}
From Table  \ref{Trend_Setter_Retweet} we observed that only 4 out of the top 20 influential authors were verified accounts. The 4 verified accounts represent an urban fashion magazine, a fashion brand, an online travel magazine, and a Chinese celebrity. The  other 16 influential authors are unverified accounts. They all seem to have a strong focus on collecting user-contributed jokes, movie trivia, quizzes, stories and so on. When we further inspected these accounts, we discovered that these accounts seem to operate as discussion and sharing platforms. The users who follow these accounts tend to contribute jokes or stories. Once they are posted, other followers tend to retweet them frequently.
\begin{table}
\centering
\begin{tabular}{|c|c|c|c|}
\hline
Author & Retweets & Topics & Retweet-Ratio\\
\hline
vovo\_panico & 11688 & 65 & 179.81 \\
cnnbrk & 8444 & 84 & 100.52 \\
keshasuja & 5110 & 51 & 100.19\\
LadyGonga & 4580 & 54 & 84.81\\
BreakingNews & 8406 & 100 & 84.06\\
MLB & 3866 & 62 & 62.35\\
nytimes & 2960 & 59 & 50.17\\
HerbertFromFG & 2693 & 58 & 46.43\\
espn & 2371 & 66 & 35.92\\
globovision & 2668 & 75 & 35.57\\
huffingtonpost & 2135 & 63 & 33.88\\
skynewsbreak & 1664 & 52 & 32\\
el\_pais & 1623 & 52 & 31.21\\
stcom & 1255 & 51 & 24.60\\
la\_patilla & 1273 & 65 & 19.58\\
reuters & 957 & 57 & 16.78\\
WashingtonPost & 929 & 60 & 15.48\\
bbcworld & 832 & 59 & 14.10\\
CBSnews & 547 & 56 & 9.76\\
TelegraphNews & 464 & 79 & 5.87\\
tweetmeme & 342 & 97 & 3.52\\
nydailynews & 173 & 51 & 3.39\\
\hline
\end{tabular}
\caption{Top Retweeted Users on Twitter contributing to at least 50 trending topics each}
\label{Twitter_retweets}
\end{table}

\begin{figure} [ht]
\centering
\includegraphics[width=80mm, height=100mm] {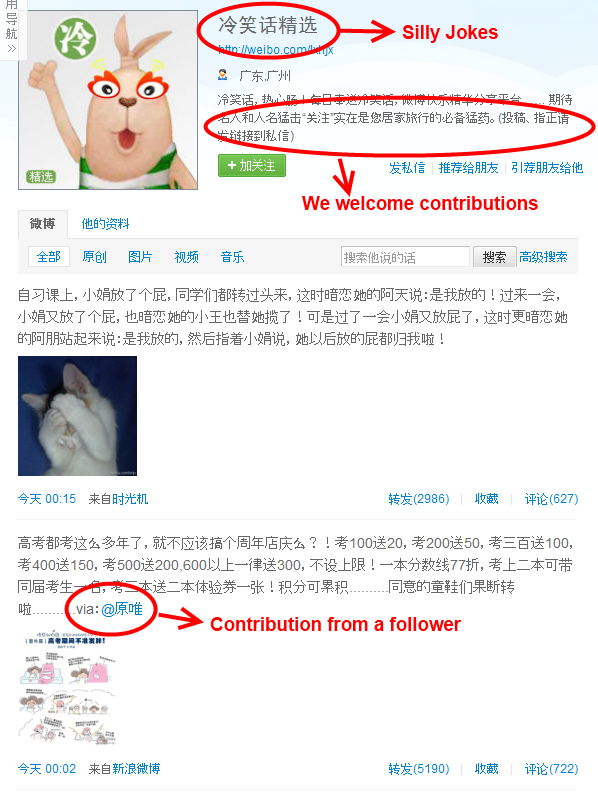}
\caption{ An Illustration of an User Account on Sina Weibo } \label{sillyjokes}
\end{figure}

We further inspect one of the accounts: ID 1644395354 (see Figure \ref{sillyjokes}); This account focuses on posting tweets about  jokes. We see from the description of the user that this account welcomes submission from followers. Followers of this account can email jokes to the account and the account administrator will post them. From Figure \ref{sillyjokes} we also see a contribution from a follower. 

The corresponding most retweeted users for trending topics on Twitter is shown in Table~\ref{Twitter_retweets}. In this case, the list is dominated by popular news sources such as CNN, the New York Times and ESPN. A large percentage of the topics that trended accordingly dealt with events in the news. This indicates that Twitter users are more attuned to news events than Sina Weibo users and amplify such articles through the medium of Twitter.
On the other hand, the consistent trend-setters on Sina Weibo are not media organizations. Instead they are unverified accounts acting as discussion forums and a platform for users to share funny pictures, jokes, and stories. We observe that none of the unverified accounts in Table \ref{Trend_Setter_Retweet} are personal accounts, with only 1 out of 4 verified accounts (in the top 20) belonging to a media organization. This represents an important contrast in the use of these media, with Chinese users being more inclined to share and propagate trivial content than the Twitter users.
\begin{table*}
\caption{Profile Information for Top 20 Retweeted Users} \label{Image}
\centering 
\begin{tabular}{|c|c|c|c|c|c|c|}
	\hline
	&	Images(\%)	&	Videos(\%)	&	Links(\%)	&	Followees	&	Followers	&	Tweets	\\
	\hline
1&	70\%	&	0\%	&	32\%		&	673	&	461398	&	719	\\
2&	57\%	&	71\%	&	0\%	&	715	&	300358	&	2508	\\
3&	21\%	&	0\%	&	17\%		&	67	&	597063	&	2600	\\
4&	30\%	&	0\%	&	0\%	&		20	&	245026	&	518	\\
5&	20\%	&	0\%	&	0\%	&		81	&	1884896	&	4261	\\
6&	8\%	&	0\%	&	0\%	&		650	&	3536888	&	10598	\\
7&	15\%	&	11\%	&	0\%		&	12	&	625117	&	804	\\
8&	46\%	&	0\%	&	0\%		&	368	&	2338610	&	3605	\\
9&	0\%	&	22\%	&	0\%		&	79	&	405847	&	716	\\
10&	5\%	&	1\%	&	0\%		&	11	&	2411888	&	17818	\\
11&	14\%	&	4\%	&	0\%		&	634	&	1551438	&	4899	\\
12&	44\%	&	16\%	&	0\%		&	352	&	6107858	&	1850	\\
13&	0\%	&	50\%	&	0\%		&	1136	&	590099	&	2041	\\
14&	6\%	&	7\%	&	0\%		&	303	&	1210833	&	11411	\\
15&	10\%	&	0\%	&	1\%		&	555	&	1220027	&	4249	\\
16&	45\%	&	0\%	&	0\%		&	13	&	615461	&	1254	\\
17&	5\%	&	40\%	&	0\%		&	12	&	496171	&	571	\\
18&	30\%	&	0\%	&	0\%		&	60	&	901612	&	3506	\\
19&	15\%	&	3\%	&	0\%		&	9	&	763264	&	2718	\\
20&	25\%	&	0\%	&	0\%		&	4	&	853877	&	2362	\\
	
	\hline
\end{tabular}
\end{table*}

\begin{table}
\centering
\begin{tabular}{|c|c|c|}
\hline
Author & Followees & Followers\\
\hline
vovo\_panico & 1069 & 154589\\
cnnbrk & 41 & 4380908\\
keshasuja & 0 & 88\\
LadyGonga & 37 & 136433\\
BreakingNews & 382 & 2570662\\
MLB & 18829 & 1237615\\
nytimes & 465 & 3250977\\
HerbertFromFG & 763 & 23318\\
espn & 286 & 1326168\\
globovision & 3582 & 753440\\
huffingtonpost & 4684 & 1042330\\
skynewsbreak & 5 & 198349\\
el\_pais & 46226 & 572260\\
stcom & 12 & 59763\\
la\_patilla & 51 & 306965\\
reuters & 603 & 724204\\
WashingtonPost & 284 & 458721\\
bbcworld & 20 & 796009\\
CBSnews & 122 & 1716649\\
TelegraphNews & 238 & 38599\\
\hline
\end{tabular}
\caption{Follower/Followee relationships for Top Retweeted Twitter Users}
\label{Twitter_folls}
\end{table}

\subsubsection{Retweets} 
When we consider the ratio of retweets, we once again observe a strong contrast with Twitter. The number of retweets that authors get on Sina Weibo are several orders of magnitude greater than the retweets for the trending topics on Twitter, although they contribute to fewer topics. This implies that the topics are trending mainly because of some content that has been retweeted many times. The $Tweets$ column in Table~\ref{Trend_Setter_Retweet} gives the unique tweets that have been retweeted. We can observe that the rate at which they have been retweeted is phenomenal. For example, the top retweeted user posted 37 tweets which were totally retweeted 1194999 times. The overall retweet percentage was around 62\% for the trending topics. In contrast, for Twitter trends, the retweets form only 31\% of the overall tweets for the trending topics. While retweets do contribute to making a topic trend on Twitter, their effect is not so large.

\subsubsection{Embedded Images, Videos and Links in Tweets}
For the list of trend-setters on Sina Weibo (Table \ref{Trend_Setter_Retweet}), we also examine the content of their tweets that appeared in the trending topics. Table \ref{Image} illustrates the percentage of these users' tweets which included images, videos, and links. We observe that a large percentage of the tweets of these users include an embedded image, and many tweets included an embedded video or a link. On the other hand, Twitter users post links in only 17.6\% of the tweets on trending topics. This is again demonstrative of the type of content that is shared in these two social media services. 

\subsubsection{Follower Relationships}
For each of the influential authors, we looked at the number of followers and followees they have, and the total number tweets they have made since their accounts are activated (Table  \ref{Image}). We discovered that most of the influential authors have more followers than followees. We hypothesize that these influential authors do not actively seek out accounts to follow. It is their content which attracts other users to follow them. Interestingly, we found this to be true on Twitter as well (shown in Table~\ref{Twitter_folls}), with the top retweeted users having very skewed follower/followee ratios. 

 \begin{figure*} [ht]
\centering
\includegraphics[width=170mm, height=70mm] {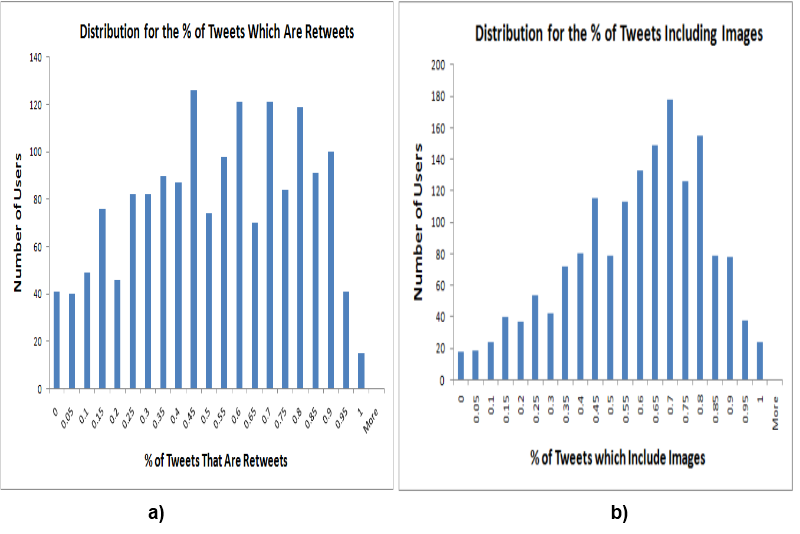}
\caption{ The Distribution for the Percentage of Tweets that are a) Retweets and b) include images} \label{percentage1}
\end{figure*} 

 \begin{figure*} 
\centering
\includegraphics[width=170mm, height=70mm] {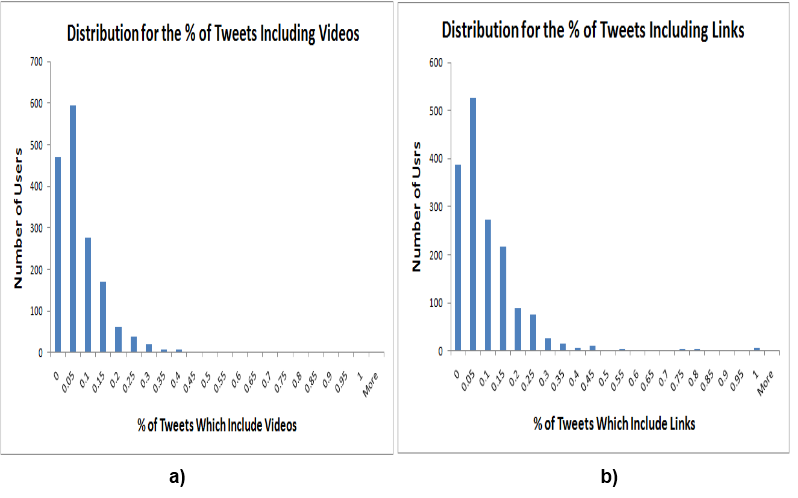}
\caption{ The Distribution for the Percentage of Tweets that include a) Videos and b) Links} \label{percentage2}
\end{figure*} 

\begin{table}
\centering
\begin{tabular}{|c|c|c|}
\hline
Rank & ID & Description\\
\hline
1 & 1757128873 & Fashion Web Magazine\\
2 & 1643830957 & Fashion Brand\\
2 & 1670645393 & Travel Web Magazine\\	
12 & 1195230310 & Celebrity\\
21 & 1740006601 & Celebrity\\
25 & 1730380283 & Game Discussion Forum\\ 	
26 & 1760945071 & Chinese Groupon\\
42 & 1322920531 & Celebrity\\
43 & 1771665380 & Record Label\\
46 & 1266321801 & Celebrity\\
48 & 1883881851 & Organization (NBA China)\\
58 & 1698229264 & Music Web Magazine\\
62 & 1642591402 & Sina Entertainment\\	
70 & 1743374541 & Pictures Discussion Forum \\
71 & 1618051664 & Sina News	\\
74 & 1653689003 & Newspaper\\
75 & 1640601392 & Sina Video	\\
82 & 1195031270 & Celebrity\\
83 & 1835254597 & Music Web Magazine\\
84 & 1830442653 & Music Web Magazine\\	
95 & 1765148101 & Sina Fashion\\
96 & 1258256457 & Celebrity \\
100 & 1596329427 & Celebrity\\	
\hline
\end{tabular}
\caption{Verified Accounts Among Top 100 Trend-setters}
\label{Verified}
\end{table}
			
\subsubsection{Verified Accounts}
When we considered the top 100 trend-setters, we found that only 23 were verified accounts. The descriptions of these 23 are shown in Table~\ref{Verified}. We observed the verified accounts are of celebrities, newspapers, magazines and a few other media sources.

% This does not indicate any Government influence in the trending topics of Sina Weibo.

\subsection{Random Profile Analysis}
From the above analyses, we observe that the Sina Weibo user accounts whose tweets are retweeted frequently and appear in multiple trending topics over time are mostly established for discussion and image/video sharing purposes. Followers use these accounts to share interesting stories, jokes, and pictures. We hypothesize that in general, users of Sina Weibo tend to retweet information more often than Twitter users. We also hypothesize that a high percentage of these users' tweets tend to include an image or a video for illustration purposes. 

To verify the above hypothesis, we selected 1732 random users on Sina Weibo. For each user, we retrieved his/her last 100 tweets and analyzed the percentage of the tweets which include images, videos, links and the percentage of tweets that are retweets. 

{\bf Retweets:} We find that on average 50.24\% of the tweets are retweets (49,76\% of the tweets are original tweets), 56.43\% of the tweets include an embedded image (43.57\% of the tweets do not), only 5.57\% of the tweets include an embedded video (94.43\% did not) and 8.03\% of the tweets include a link (91.94\% did not). We observe that over half of the tweets by our 1732 random users are retweets, and over half of the tweets include an embedded image. 
Figure \ref{percentage1} a) illustrates the histogram for the number of users with a certain percentage of tweets which are retweets. We see that the distributions for the number of users are fairly even throughout all the percentages and is especially high around 45\%, 60\%, 70\% and 80\%. 

{\bf Content:} Figure \ref{percentage1} b) illustrates the histogram for the number of users with a certain percentage of tweets which include an embedded image. We see that the distribution for the number of users peak at around 55\% and 80\% while remain low below 40\% and above 85\%. Figure \ref{percentage2} a) and \ref{percentage2} b) illustrates the histograms for the number of users with a certain percentage of tweets which include a video or a link. We see that both distributions  peaked at 5\% and then diminish quickly after.
In the case of twitter, all forms of media are shared through the use of hyperlinks. It has been shown in \cite{Romero2011} that URLs form 1/15th (6.6\%) of all twitter traffic. This is a very low percentage when compared to Sina Weibo.

\section{Conclusion and Future Work}

We analyzed the tweets that are responsible for causing trending topics on Sina Weibo  and the users that created these tweets.  We observed that there are vast differences between the content that is shared on Sina Weibo than that of Twitter. People tend to use Sina Weibo to share jokes, images and videos and a significantly large percentage of posts are retweets. The trends that are formed are almost entirely due to the repeated retweets of such media content.  In contrast, we observed on Twitter that trending topics are mainly caused by sources of media \cite{Asur2011}. 

%We also observed that none of the verified accounts in our list of top 100 trend-setters are affiliated with any Government agencies. 
As future work, we plan to conduct further temporal analysis regarding the evolution of trends on Sina Weibo. We also plan to identify the patterns of retweets between followers of influential authors.

%
% The following two commands are all you need in the
% initial runs of your .tex file to
% produce the bibliography for the citations in your paper.
\bibliographystyle{abbrv}
\bibliography{UvicThesis}  % sigproc.bib is the name of the Bibliography in this case

\begin{thebibliography}{10}

\bibitem{Agarwal2008Identifying}
N.~Agarwal, H.~Liu, L.~Tang, and P.~S. Yu.
\newblock {Identifying the Influential Bloggers in a Community}.
\newblock {\em WSDM'08}, 2008.

\bibitem{Asur2011}
S.~Asur, B.~A. Huberman, G.~Szabo, and C.~Wang.
\newblock Trends in social media - persistence and decay.
\newblock In {\em 5th International AAAI Conference on Weblogs and Social
  Media}, 2011.

\bibitem{Backstrom}
L.~Backstrom, D.~Huttenlocher, J.~Kleinberg, and X.~Lan.
\newblock Group formation in large social networks: membership, growth, and
  evolution.
\newblock In {\em Proceedings of the 12th International Conference on Knowledge
  Discovery and Data Mining}, pages 44--54. ACM, 2006.

\bibitem{StrongTie}
Y.~Bian.
\newblock Bringing strong ties back in: indirect ties, network bridges, and job
  searches in china.
\newblock {\em American Sociological Review}, 62(3):366--385, 1997.

\bibitem{Guanxi3}
Y.~Bian, R.~Breiger, D.~Davis, and J.~Galaskiewicz.
\newblock Occupation, class, and social networks in urban china.
\newblock {\em Social Forces}, 83(4):1443--1468, 2005.

\bibitem{Buchanan}
M.~Buchanan.
\newblock {\em Nexus: Small Worlds and the Groundbreaking Theory of Networks}.
\newblock {W. W. Norton \& Company}, May 2003.

\bibitem{Carrington}
P.~J. Carrington, J.~Scott, and S.~Wasserman, editors.
\newblock {\em Models and Methods in Social Network Analysis}.
\newblock Cambrige University Press, 2005.

\bibitem{Cha2010Measuring}
M.~Cha, H.~Haddadi, F.~Benevenuto, and K.~P. Gummadi.
\newblock {Measuring User Influence in Twitter: The Million Follower Fallacy}.
\newblock In {\em Fourth International AAAI Conference on Weblogs and Social
  Media}, May 2010.

\bibitem{Statistic-general}
CNNIC.
\newblock The 21st statistics report on the internet development in china (in
  chinese), 2010.

\bibitem{Statistic-rural}
CNNIC.
\newblock Survey report on internet development in rural china (in chinese),
  2010.

\bibitem{Crandall}
D.~Crandall, D.~Cosley, D.~Huttenlocher, J.~Kleinberg, and S.~Suri.
\newblock Feedback effects between similarity and social influence in online
  communities.
\newblock In {\em Proceedings of the 14th ACM SIGKDD international conference
  on Knowledge discovery and data mining}, pages 160--168. ACM, 2008.

\bibitem{768262}
J.-L. Farh, A.~S. Tsui, K.~Xin, and B.-S. Cheng.
\newblock The influence of relational demography and guanxi: the {C}hinese
  case.
\newblock {\em Organization Science}, 9(4):471--488, 1998.

\bibitem{Goyal}
A.~Goyal, F.~Bonchi, and L.~V.~S. Lakshmanan.
\newblock Learning influence probabilities in social networks.
\newblock In {\em Web Search and Data Mining}, pages 241--250, 2010.

\bibitem{Huberman}
B.~A. Huberman, D.~M. Romero, and F.~Wu.
\newblock Social networks that matter: Twitter under the microscope.
\newblock {\em Computing Research Repository}, 2008.

\bibitem{Jamali}
M.~Jamali and H.~Abolhassani.
\newblock Different aspects of social network analysis.
\newblock In {\em Proceedings of the 2006 IEEE/WIC/ACM International Conference
  on Web Intelligence}, pages 66--72, 2006.

\bibitem{Jin}
L.~Jin.
\newblock {Chinese outline BBS sphere: what BBS has brought to China}.
\newblock Master's thesis, Massachusetts Institute of Technology, April 2009.

\bibitem{Kempe05influentialnodes}
D.~Kempe, J.~Kleinberg, and E.~Tardos.
\newblock Influential nodes in a diffusion model for social networks.
\newblock In {\em Proceedings of 32nd International Colloquium on Automata,
  Languages and Programming}, pages 1127--1138. Springer Verlag, 2005.

\bibitem{Kumar}
R.~Kumar, J.~Novak, and A.~Tomkins.
\newblock Structure and evolution of online social networks.
\newblock In {\em Proceedings of the 12th ACM SIGKDD International Conference
  on Knowledge Discovery and Data Mining}, pages 611--617. ACM, 2006.

\bibitem{Kwak}
H.~Kwak, C.~Lee, H.~Park, and S.~Moon.
\newblock What is twitter, a social network or a news media?
\newblock In {\em Proceedings of the 19th international conference on World
  wide web}, WWW '10, pages 591--600, 2010.

\bibitem{Leskovec05patternsof}
J.~Leskovec, A.~Singh, and J.~Kleinberg.
\newblock Patterns of influence in a recommendation network.
\newblock In {\em Proceedings of Pacific-Asia Conference on Knowledge Discovery
  and Data Mining}, pages 380--389. Springer-Verlag, 2005.

\bibitem{Mathioudakis}
M.~Mathioudakis and N.~Koudas.
\newblock Twittermonitor: trend detection over the twitter stream.
\newblock In {\em Proceedings of the 2010 international conference on
  Management of data}, SIGMOD '10, pages 1155--1158, 2010.

\bibitem{mcpherson2001birds}
M.~McPherson, L.~Smith-Lovin, and J.~M. Cook.
\newblock Birds of a feather: homophily in social networks.
\newblock {\em Annual Review of Sociology}, 27(1):415--444, 2001.

\bibitem{Mislove}
A.~Mislove, M.~Marcon, K.~P. Gummadi, P.~Druschel, and B.~Bhattacharjee.
\newblock Measurement and analysis of online social networks.
\newblock In {\em Proceedings of the 7th SIGCOMM Conference on Internet
  Measurement}, pages 29--42. ACM, 2007.

\bibitem{Romero2011}
D.~M. Romero, W.~Galuba, S.~Asur, and B.~A. Huberman.
\newblock Influence and passivity in social media.
\newblock In {\em 20th International World Wide Web Conference (WWW'11)}, 2011.

\bibitem{WorkControl}
D.~Ruan.
\newblock Interpersonal networks and workplace controls in urban china.
\newblock {\em The Australian Journal of Chinese Affairs}, 29:89--105, 1993.

\bibitem{Tang}
J.~Tang, J.~Sun, C.~Wang, and Z.~Yang.
\newblock Social influence analysis in large-scale networks.
\newblock In {\em Proceedings of the 15th ACM SIGKDD international conference
  on Knowledge discovery and data mining}, KDD '09, pages 807--816, 2009.

\bibitem{Netizen}
F.~Y. Wang.
\newblock Beyond x 2.0: where should we go?
\newblock {\em IEEE Intelligent Systems}, 24(3):2--4, 2009.

\bibitem{Wu2}
S.~Wu, J.~M. Hofman, W.~A. Mason, and D.~J. Watts.
\newblock Who says what to whom on twitter.
\newblock In {\em Proceedings of the 20th international conference on World
  wide web}, WWW '11, pages 705--714, 2011.

\bibitem{Xin}
M.~Xin.
\newblock Chinese bulletin board system's influence upon university students
  and ways to cope with it (in chinese).
\newblock {\em Journal of Nanjing University of Technology (Social Science
  Edition)}, 4:100--104, 2003.

\bibitem{Xu}
Y.~Xu, J.~A.~M. Farver, D.~Schwartz, and L.~Chang.
\newblock Social networks and aggressive behavior in chinese children.
\newblock {\em International Journal of Behavioral Development}, 28:401--410,
  2004.

\bibitem{Tai}
T.~Z. Xue.
\newblock {\em The Internet in China : Cyberspace and Civil Society}.
\newblock Routledge, 2006.

\bibitem{yu}
L.~Yu and V.~King.
\newblock The evolution of friendships in chinese online social networks.
\newblock {\em Social Computing / IEEE International Conference on Privacy,
  Security, Risk and Trust, 2010 IEEE International Conference on}, 0:81--87,
  2010.

\end{thebibliography}

\end{document}